# Theoretical study on excited states of $ICl^+$ molecular ion considering spin-orbit coupling*

Rui Li[a,b*], Ronglong DOU[a], Ting GAO[a], Qinan LI[a], Chaoqun SONG[a]

[a)] Department of Physics, College of Science, Qiqihar University, Qiqihar 161006, China

[b)] College of Teacher Education, Qiqihar University, Qiqihar 161006, China

**Abstract:** The electronic structure of the $ICl^+$ molecular ion is investigated by using high-level multireference configuration interaction (MRCI) method. To improve computational accuracy, Davidson corrections, spin-orbit coupling (SOC), and core-valence electron correlations effects are incorporated into the calculations. The potential energy curves (PECs) of 21 Λ-S states associated with the two lowest dissociation limits $I^+(^1D_g)+Cl(^2P_u)$ and $I^+(^3P_g)+Cl(^2P_u)$ are obtained. The dipole moments (DMs) of the 21 Λ-S states of $ICl^+$ are systematically studied, and the variations of DMs of the identical symmetry state ($2^2\Sigma^+/3^2\Sigma^+$ and $2^2\Pi/3^2\Pi$) in the avoided crossing regions are elucidated by analyzing the dominant electronic configuration. When considering the SOC effect, the Λ-S states with the same Ω components may form new avoided crossing point, making the PECs more complex. With the help of the calculated SOC matrix element, the interaction between crossing states can be elucidated. Spin-orbit coupling matrix elements involving the $2^2\Pi$, $3^2\Pi$, $1^2\Delta$ and $2^2\Delta$ states are calculated. By analyzing potential energy curves of these states

---

Corresponding authors: Email:lirei01@163.com

and the nearby electronic states, the possible predissociation channels are provided. Based on the computed PECs, the spectroscopic constants of bound Λ-S and Ω states are determined. The comparison of the spectroscopic constants between Λ-S and Ω states indicates that the SOC effect has an obvious correction to the spectroscopic properties of low-lying states. Finally, the transition properties between excited states and the ground state are studied. Based on the computed transition dipole moments and Franck-Condon Factors, radiative lifetimes for the low-lying vibrational levels of excited states are evaluated. All the data presented in this paper are openly available at https://doi.org/10.57760/sciencedb.j 00213.00140.

## 1. Introduction

The interhalogen compounds formed by VII A group have an important impact on the environment, among which the contents of I, Cl and Br elements have an important impact on the concentration of ozone ($O_3$) in the atmosphere, which is also one of the causes of global warming[1–3]. In addition, iodine has rich supramolecular chemical properties and plays an increasingly important role in the synthesis of new organic and inorganic substances and materials, as well as in the main components of electrochemical devices such as third-generation solar cells[4]. In view of the importance of iodine in many aspects such as environment and materials, more and more researchers have begun to study the halides of iodine, especially the electronic structure and spectral properties of iodine chloride molecular ion ($ICl^+$).

In the experimental study, Evans and Orchard[5] observed the photoelectron spectrum of ICl molecule with a resolution of 50 meV. The results show that the spin-orbit component of the dissociation of the $X^2\Pi_i$ band system of the $ICl^+$ molecular ion has an obvious vibrational structure. Potts and Price[6] reported the high-resolution photoelectron spectrum of ICl molecule, and explained the electronic structure of ICl molecule. The vertical and adiabatic ionization potentials and emission spectra of the ionization state produced by removing one electron from each molecular orbital were given. Subsequently, Eland[7] used photoelectron-photoion coincidence technique to study the angular distributions, energy disposal and branching studied of $ICl^+$ molecular ion, and found that some levels of $A^2\Pi_i$ state of $ICl^+$ molecular ion have dissociation stability. Dibeler et al.[8] recorded ion yield curves in the range of 1000-1250 Å, but did not analyze the spectra. Venkateswarlu[9] observed the vacuum ultraviolet spectrum of Rydberg states of ICl molecule, and pointed out that the energy of 81362±80 $cm^{-1}$ in the spectral band produced by the transition of Rydberg states may correspond to the $X^2\Pi_{3/2}$ state of the $ICl^+$ molecular ion. In another study, Tuckett et al.[10] did not observe the fluorescence of $ICl^+$ molecular ion by using photoion-fluorescence photon coincidences technique, and obtained the electron branching ratio of $A^2\Pi$ state of 0.4 by using glass/teflon inlet system and NeI excitation for $ICl^+$ molecular ion. Kaur et al.[11] studied the formation of ionic pairs in the ICl molecule. In the initial ionization region, it was found that the heights of the positive and negative ion curves were basically the same, confirming their joint role in the dissociation of ionic pairs. Yencha et al.[12] conducted a high-resolution (1–11 meV) threshold photoelectron spectroscopy study on $ICl^+$ molecular ions in the molecular valence ionization region using synchrotron radiation and a penetrating-field electron spectrometer. They obtained precise vibrational constants for the two spin–orbit components of the $ICl^+$ molecular ion in the $X^2\Pi$ state. They found that the adiabatic ionization potentials for the $ICl^+$ molecular ion in the $X^2\Pi_{3/2}$ and $X^2\Pi_{1/2}$ states at $v^+$=0 are 10.076±0.002 eV and 10.655±0.002 eV, respectively, and determined the adiabatic ionization potential for the $A^2\Pi$ state at $v^+$=0 to be 12.5 eV. Ridley et al.[13] reanalyzed the vacuum ultraviolet absorption spectrum of ICl molecules using new data from zero kinetic energy-pulsed field ionization (ZEKE-PFI) photoelectron spectra of ICl molecules and (2+1) resonance enhanced multiphoton ionization (REMPI) spectrum of ICl molecules, obtaining the adiabatic ionization potential for the ground state $X^2\Pi_{3/2}$ ($v^+$=2) of the $ICl^+$ molecular ion as 81246±3 $cm^{-1}$.

Compared with the experimental studies, there are relatively few theoretical studies on the electronic structure of $ICl^+$ molecular ions so far. Straub and McLean[14] have calculated the electronic structures of ICl molecule and the molecular ion of $ICl^+$ by using the self-consistent field molecular orbital method. The ionization potential of ICl molecule is given. Subsequently, Dyke et al.[15] used the Hartree-Fock method with relativistic correction to give the vertical ionization energies of a series of electronic states of $ICl^+$ molecular ions. Balasubramanian[16] has calculated the potential energy curves (PECs) of the lower energy electronic states of $ICl^+$ molecular ion by using the relativistic configuration interaction (RCI) method, and obtained the spectroscopic constants of the four bound states $X^2\Pi_{3/2}$, $X^2\Pi_{1/2}$,$^2\Pi$(II) and $^2\Sigma^+_{1/2}$. The configuration components of the lower excited states of $ICl^+$ molecular ion have been discussed. At present, only Balasubramanian[16] has studied the electronic structure and spectroscopic constants of several low energy electronic states of $ICl^+$ molecular ion by using the configuration interaction method with relativistic effects. Previous theoretical studies have not elucidated the interaction between electronic states, the predissociation mechanism, and the radiative transition properties. In this work, the energy eigenvalues of 21 Λ-S states and 42 Ω states are calculated by using the high-level multi-reference configuration interaction (MRCI) method. The electronic structure, spectroscopic constants and transition properties of $ICl^+$ molecular ion have been studied. In addition, the predissociation mechanism of the low-lying excited state is studied, and the radiative lifetime of the low-lying excited bound state is predicted.

## 2. Method and calculation details

*Ab initio* calculations for the low-lying excited states of the molecular ion of $ICl^+$ were performed with the MOLPRO[17] program. In the calculation, the basis set used for I atom is aug-cc-pwCVQZ-PP basis set including pseudopotential ECP28MDF, and the basis set used for Cl atom is aug-cc-pVQZ basis set[18,19]. All calculations in this work are performed under the Abelian subgroup $C_{2v}$ point group symmetry of the $C_{\infty v}$ point group. The $C_{2v}$ point group contains 4 irreducible representations as $A_1$, $B_1$, $B_2$ and $A_2$. The correspondences between the spectral terms of the molecule and the irreducible representations of the $C_{2v}$ point group are as follows: $\Sigma^+ \rightarrow A_1$, $\Sigma^- \rightarrow A_2$, $\Pi \rightarrow B_1+B_2$, $\Delta \rightarrow A_1+A_2$ and $\Phi \rightarrow B_1+B_2$.

The specific calculations for this molecular structure were carried out through the following steps. Based on the single-configuration wave function of the ground state of the $ICl^+$ molecular ion obtained using the Hartree-Fock (HF) self-consistent field method, the multi-configuration wave function was calculated using the complete active space self-consistent field (CASSCF) method [20,21]. Subsequently, the optimized CASSCF wave function was used to perform MRCI calculations [22,23]. To account for the effect of the truncated configuration space on the accuracy of the electronic structure, the Davidson correction (+*Q*) was introduced[24]. In CASSCF calculation, the selection of active space is very important. After testing, 11 molecular orbitals $5a_1$, $3b_1$ and $3b_2$ are selected as the active space of the molecule. In the MRCI+$Q$ calculation, the electrons in the 4s4p (I) shell and the 1s2s (Cl) shell are put into the frozen core orbitals, and the 4d orbitals of I are included in the core-valence electron correlation energy calculation as five inactive orbitals. The spin-orbit coupling (SOC) is used as a perturbation to diagonalize the spin-orbit Hamiltonian matrix[25] using the spin-orbit ECP operator, and the eigenenergies and eigenfunctions incorporating the SOC effect are obtained, and the PECs of the Ω state are given. Based on the calculated PECs, the one-dimensional nuclear Schrödinger equation is solved by the LEVEL program[26] to obtain the spectroscopic constants of Λ-S and Ω states.

## 3. Discussion of results

### 3.1 Potential energy curve, spectroscopic constants and dipole moment of the Λ-S state of the $ICl^+$ molecular ion

The eigenenergies of 21 Λ-S states corresponding to the two lowest dissociation limits $I^+(^3P_g)+Cl(^2P_u)$, $I^+(^1D_g)+Cl(^2P_u)$ of $ICl^+$ molecular ion are calculated by using the high-level MRCI method. Previous theoretical calculations show that the Davidson correction and the core-valence electron correlation energy have a significant effect on the shape of the potential energy curve of the excited state, so we consider the Davidson correction and the core-valence electron correlation energy in the MRCI calculation of $ICl^+$ molecular ion[27–32]. Figure1. gives 15 doublet and 6 quartet PECs. Table 1 lists the main electronic configurations, their percentages and vertical excitation energies of the ground and excited states of the $ICl^+$ molecular ion at the equilibrium internuclear distance ($R_e$=2.25 Å), and the spectroscopic constants of the 10 bound states are listed in Table 2. It can be seen from Table 1 that the electronic configuration components of the excited states $1^4\Sigma^-$, $2^2\Pi$, $1^2\Delta$, $1^2\Sigma^+$, $3^2\Pi$, $3^2\Sigma^-$ all show obvious multi-configuration characteristics, and the accuracy of the excited state electronic structure can be ensured by using the MRCI method.

The ground state of the $ICl^+$ molecular ion is $X^2\Pi$, and its main electronic configuration components is $7\sigma^2 8\sigma^2 9\sigma^2 10\sigma^2 11\sigma^0 3\pi^4 4\pi^4 5\pi^3$. The calculated value of the ground state $D_0$ at MRCI level is 2.50 eV, which is in good agreement with the experimental result of 2.52 eV[33]. The main electronic configurations of the $2^2\Pi$ state are $7\sigma^2 8\sigma^2 9\sigma^2 10\sigma^2 11\sigma^0 3\pi^4 4\pi^3 5\pi^4$(87%) and $7\sigma^2 8\sigma^2 9\sigma^2 10\sigma^1 11\sigma^1 3\pi^4 4\pi^4 5\pi^3$(6%), which correspond to the 4π→5π transition and 10σ→11σ transition of the ground state, respectively. The harmonic vibrational frequency $\omega_e$ and the equilibrium internuclear distance $R_e$ of $2^2\Pi$ are 259 $cm^{-1}$ and 2.59 Å, respectively, which are consistent with the results of 207 $cm^{-1}$ and 2.78 Å calculated by Balasubramanian[16] using relativistic configuration interaction. The main electronic configurations of $3^2\Pi$ are $7\sigma^2 8\sigma^2 9\sigma^2 10\sigma^2 11\sigma^2 3\pi^4 4\pi^4 5\pi^1$(48%) and $7\sigma^2 8\sigma^2 9\sigma^2 10\sigma^2 11\sigma^2 3\pi^4 4\pi^3 5\pi^2$(48%). The calculated adiabatic excitation energy $T_e$ of the $3^2\Pi$ state is 22122 $cm^{-1}$, which is only 298 $cm^{-1}$ (1.3%) smaller than the experimental result of 22420 $cm^{-1}$[33]. The main electronic configuration of $1^4\Delta$, $3^2\Sigma^+$,$2^2\Delta$ state is $7\sigma^2 8\sigma^2 9\sigma^2 10\sigma^2 11\sigma^1 3\pi^4 4\pi^3 5\pi^3$, that is, one electron in the ground state of 4π is excited to the 11σ state. The main electronic configuration of the $2^2\Sigma^+$ state is $7\sigma^2 8\sigma^2 9\sigma^2 10\sigma^1 11\sigma^0 3\pi^4 4\pi^4 5\pi^4$(94%), which excites an electron from the ground state 10σ to the outer 5π.

The bonding characteristics of molecules can be reflected by dipole moments (DMs). Therefore, in this paper, the dipole moments of 21 Λ-S states are calculated at the MRCI level, and the results are listed in Figure 2. For intuition, the dipole moment curves of the doublet and quartet Λ-S states are given Figure 2(a),(b), respectively. As shown in the Figure 2, the dipole moment curves of all electronic states gradually tend to positive infinity with the increase of the internuclear distance, indicating that the dissociation products of the lowest two dissociation limits are ions. From the Figure 1., it can be seen that $2^2\Pi/3^2\Pi$ and $2^2\Sigma^+/3^2\Sigma^+$ exhibit obvious avoided crossing phenomena at internuclear distances of 3.05-3.20 Å and 2.50-2.65 Å, respectively. At the same time, it can be found from Figure 2 that the DMs curves of $2^2\Pi/3^2\Pi$ and $2^2\Sigma^+/3^2\Sigma^+$ also change significantly in the above region. To explain this phenomenon more clearly, Figure 3. gives the variation of the correlation weights of $2^2\Pi/3^2\Pi$ and $2^2\Sigma^+/3^2\Sigma^+$ with the internuclear distance.

The Figure 3(a) shows the variation of the main electron configuration components con-A and con-B for the $2^2\Pi$ state (solid line) and the $3^2\Pi$ state (dashed line). Where con-A is $7\sigma^2 8\sigma^2 9\sigma^2 10\sigma^2 11\sigma^2 3\pi^4 4\pi^3 5\pi^2$ and con-B is $7\sigma^2 8\sigma^2 9\sigma^2 10\sigma^2 11\sigma^0 3\pi^4 4\pi^3 5\pi^4$. When the internuclear distance is small, con-A in the $3^2\Pi$ state plays a key role, but the role weakens rapidly when the avoided crossing point ($R_{ACP}$) is approached; Its main electronic component, con-B, begins to play an important role. The main electronic configurations of $2^2\Pi$ and $3^2\Pi$ states show complementary trends at the boundary point of $R_{ACP}$.

The Figure 3(b) shows the variation of the main electronic configuration components con-C and con-D for the $3^2\Sigma^+$ state (solid line) and the $2^2\Sigma^+$ state (dashed line). Con-C is $7\sigma^2 8\sigma^2 9\sigma^2 10\sigma^1 11\sigma^0 3\pi^4 4\pi^4 5\pi^4$, and con-D is $7\sigma^2 8\sigma^2 9\sigma^2 10\sigma^2 11\sigma^1 3\pi^4 4\pi^3 5\pi^3$. When the internuclear distance is small, con-C plays an important role in the $2^2\Sigma^+$ state. Also after the $R_{ACP}$ point, the main electronic configuration component of the $2^2\Sigma^+$ state changes from con-C to con-D. The main configuration composition changes of $2^2\Sigma^+$ and $3^2\Sigma^+$ states also show complementary characteristics with $R_{ACP}$ as the demarcation point. The main configuration components of $2^2\Pi/3^2\Pi$ and $2^2\Sigma^+/3^2\Sigma^+$ are exchanged in the $R_{ACP}$ region. This phenomenon fully shows that the configuration components of electronic states with the same symmetry will be interchanged after and $R_{ACP}$. Moreover, the change of the main configuration components of $2^2\Pi/3^2\Pi$ and $2^2\Sigma^+/3^2\Sigma^+$ can lead to the mutation of the DMs in the $R_{ACP}$ region.

### 3.2 Spin-orbit coupling and predissociation of the Λ-S state of $ICl^+$ molecular ion

In order to study the interaction between the electronic states caused by the curve crossing of the Λ-S states and the predissociation caused by the coupling of these electronic states, the Figure 4. shows a magnified view of the four crossing regions, and the SOC matrix elements including the relevant electronic states in the crossing region are calculated. The two crossover regions are located within the energy range of 17 000-22 000 $cm^{-1}$ in Figure 4(a). The first crossover region is located at $R$=2.60-2.80 Å and is caused by the crossover of the $2^2\Pi$ state and the $1^2\Sigma^-$, $1^2\Sigma^+$, $1^4\Sigma^+$,$1^2\Delta$, and $1^4\Delta$ states. The second crossover region is at *R*=2.80-2.95 Å and is caused by the crossover of the $1^2\Delta$ state and the $1^4\Sigma^+$, $1^4\Delta$ states. The two crossing regions in Figure 4(b) are located in the energy range of 22000-26000 cm$^{-1}$. The third crossover region is at *R*=2.70-2.95 Å, which is caused by the crossover of $2^2\Delta$ state and $1^4\Pi$, $2^4\Pi$, $3^2\Pi$ states. The fourth crossover region is at *R*=3.00-3.20 Å, which is caused by the crossover of $3^2\Pi$ state and $1^4\Pi$, $2^4\Pi$ states. In order to further study the interaction of these crossing states, the calculated SOC matrix elements involving the $2^2\Pi/1^2\Delta$ and $2^2\Delta/3^2\Pi$ states are given by Figure 5(a),(b). As shown in Figure 4(a), near the vibrational levels $v'$=0, $v'$=1 and $v'$=2, the $2^2\Pi$ state and the $1^2\Sigma^-$,$1^2\Sigma^+$,$1^4\Sigma^+$,$1^2\Delta$, $1^4\Delta$ states cross. The absolute values of the SOC matrix elements at the crossing points all exceed 664 cm$^{-1}$, which leads to the following predissociation of the $2^2\Pi$ state: $2^2\Pi(v'\geq2)\rightarrow1^2\Sigma^-$, $2^2\Pi(v'\geq1)\rightarrow1^2\Sigma^+$, $2^2\Pi(v'>1)\rightarrow1^4\Sigma^+$, $2^2\Pi(v'\geq0)\rightarrow1^2\Delta$, $2^2\Pi(v'\geq1)\rightarrow1^4\Delta$. The $1^2\Delta$ state intersects with the $1^4\Sigma^+$,$1^4\Delta$ states near the vibrational level $v'$=0, and the absolute values of the SOC matrix elements at the intersection point exceed 1484 cm$^{-1}$, which satisfies the condition for predissociation. The $1^2\Delta$ state $v'\geq0$ vibrational level is predissociated through the $1^2\Delta\rightarrow1^4\Sigma^+$, $1^2\Delta\rightarrow1^4\Delta$ channels. Near the vibrational levels $v'$=0 and $v'$=1, the $2^2\Delta$ state intersects with $1^4\Pi$, $2^4\Pi$, and $3^2\Pi$, and the absolute values of the SOC matrix elements at the intersection points all exceed 454 cm$^{-1}$, which leads to the predissociation of the vibrational levels of the $2^2\Delta$ state through the channels $2^2\Delta(v'\geq1)\rightarrow1^4\Pi$, $2^2\Delta(v'\geq0)\rightarrow2^4\Pi$, $2^2\Delta(v'\geq0)\rightarrow3^2\Pi$. The $3^2\Pi$ state intersects with the $1^4\Pi$ and $2^4\Pi$ states near the vibrational levels $v'$=0,$v'$=1 and $v'$=2, and the absolute values of the SOC matrix elements at the intersection points all exceed 321 cm$^{-1}$, which also meets the condition for predissociation. The $3^2\Pi$ state $v'\geq0$ vibrational level is predissociated through the $3^2\Pi\rightarrow1^4\Pi$ and $3^2\Pi\rightarrow2^4\Pi$ channels.

### 3.3 Potential energy curves and spectroscopic constants for Ω states of $ICl^+$ molecular ion

The SOC effect affects the structure of the electronic states, causing the avoided crossing of Λ-S states with the same Ω component, which makes the PECs in these avoided crossing regions extremely complex. The energy of the Ω states is obtained by solving the eigenvalue equation for the Hamiltonian of the molecular system containing the ECP operator. The Table 3 lists the energies between the dissociation relation and the dissociation limit for these Ω states. The energy differences of other dissociation limits relative to the lowest dissociation limit $I(^3Pg_2)+Cl(^2Pu_{3/2})$ are 821, 6433, 6901, 7254, 7722, 13950 and 14771 $cm^{-1}$, and the differences from the experimental results[34] are 61 $cm^{-1}$ (6.91%), 15 $cm^{-1}$ (0.23%), 186 $cm^{-1}$ (2.62%), 76 $cm^{-1}$ (1.04%), 247 $cm^{-1}$ (3.10%), 223 $cm^{-1}$ (1.62%) and 161 $cm^{-1}$ (1.10%). In addition, the calculated SOC splitting of $Cl(^2P_{3/2})$-$Cl(^2P_{1/2})$ is 821 $cm^{-1}$, which is in good agreement with the experimental measurement of 882 $cm^{-1}$[34]. The SOC effect leads to the splitting of 21 Λ-S states of the $ICl^+$ molecular ion into 42 Ω states. For clarity, PECs with the same symmetry Ω=1/2, Ω=3/2, Ω=5/2 and Ω=7/2 are shown in Figure 6(a)-(d), respectively. Based on the theoretically calculated potential energy curves of the Ω states, the spectroscopic constants of the bound Ω states were obtained by numerical integration Table 4. Under the influence of SOC effect, the ground state $X^2\Pi$ is split into two Ω states $X^2\Pi_{1/2}$ and $X^2\Pi_{3/2}$, the quartet $1^4\Sigma^-$ is split into two Ω states 1/2(II) and 3/2(II) with the splitting of 394 $cm^{-1}$.

Taking into account the SOC effect, our calculated $\omega_e$ of the ground state $X^2\Pi_{3/2}$ is 419 $cm^{-1}$, which is 29 $cm^{-1}$ (7.42%) higher than the experimental result of 390 $cm^{-1}$[6]. It is only 10 $cm^{-1}$ (2.33%) smaller than the $\omega_e$ value of 429 $cm^{-1}$ observed by Yencha et al.[12]. Our calculated $R_e$ of $X^2\Pi_{3/2}$ is 2.250 Å, which is consistent with the experimental result of 2.240±0.01 Å[12], and the deviation is significantly reduced compared with the previous theoretical result of 2.470 Å[16]. For $X^2\Pi_{1/2}$, $\omega_e$ and $\omega_e x_e$, the calculated results are 426 $cm^{-1}$ and 2.13 $cm^{-1}$, respectively, which are only 11 $cm^{-1}$ (2.52%) and 0.05 $cm^{-1}$ (2.40%) different from the experimental values of 437 $cm^{-1}$ and 2.08 $cm^{-1}$ of Yencha et al.[12]. Our calculated $R_e$ of $X^2\Pi_{1/2}$ is 2.248 Å, which is only 0.024 Å (1.08%) higher than the experimental result 2.224±0.001 Å[12], and the error is significantly reduced compared with the previous theoretical result 2.460 Å[16]. Our calculated spin-orbit splitting of $X^2\Pi_{1/2}$-$X^2\Pi_{3/2}$ is 4501 $cm^{-1}$, which is only 179 $cm^{-1}$ (3.82%)[33] and 169 $cm^{-1}$ (3.62%) smaller than the experimental values of 4680 $cm^{-1}$ and 4670±16 $cm^{-1}$[12].

Compared with the $1^4\Sigma^-$ of the pure Λ-S state, the calculated $T_e$ of 1/2(II) and 3/2(II) are increased by 1191 and 1585 $cm^{-1}$. Considering the influence of SOC effect, the $\omega_e$ value of $1^4\Sigma^-_{1/2}$(1/2(II)) changes from 218 $cm^{-1}$ to 244 $cm^{-1}$, increasing by 26 $cm^{-1}$; the $\omega_e$ value of $1^4\Sigma^-_{3/2}$(3/2(II)) changes to 156 $cm^{-1}$, decreasing by 62 $cm^{-1}$. The $D_0$ of 1/2(II) and 3/2(II) obtained by theoretical calculation are 0.50 and 0.44 eV, respectively, which is significantly smaller than the result of 0.83 eV for the pure Λ-S state $1^4\Sigma^-$. In summary, considering the obvious changes of $T_e$ and $D_0$, it can be found that the SOC effect has a significant influence on the shape of PECs and the dissociation energy $D_0$ of the excited states in the Franck-Condon region.

### 3.4 Transition properties of $ICl^+$ molecular ion

According to the Franck-Condon principle, the Franck-Condon factor can represent the intensity distribution of the *v′-v″* transition spectral band. The transition dipole moments (TDMs) are calculated at the MRCI level, and the transition properties of the transition dipole moments molecular ion are clarified. The calculated TDMs for $2^2\Sigma^+$-$X^2\Pi$, $3^2\Sigma^+$-$X^2\Pi$, $1^2\Delta$-$X^2\Pi$ and $2^2\Delta$-$X^2\Pi$ are shown in Figure 7(a). With the increase of internuclear distance, the TDM of $2^2\Delta$-$X^2\Pi$ transition decreases gradually, and the transition curve of $1^2\Delta$-$X^2\Pi$ decreases gradually between the internuclear distance of 2. 00 and 2. 50 Å, then increases and decreases again, and a peak with a peak value of 0. 018 a.u. Appears at the internuclear distance of 2. 90 Å. The transition curves of $2^2\Sigma^+$-$X^2\Pi$ and $3^2\Sigma^+$-$X^2\Pi$ at the internuclear distance of 2.50-2.75 Å are changed, which is due to the avoided crossing phenomenon between the electronic states with the same symmetry. The TDMs curves of 1/2(II)-$X^2\Pi_{3/2}$, 1/2(III)-$X^2\Pi_{3/2}$ and 3/2(III)-$X^2\Pi_{3/2}$ are given by Figure 7(b). As shown in Figure 7, all TDMs approach zero as the bond length increases beyond the Franck-Condon regime. 1/2(II) the main Λ-S component is $1^4\Sigma^-$(79.7%), 3/2(III) the main Λ-S components are $1^4\Delta$(58.3%) and $1^2\Delta$(34.8%), and 1/2(III) the main Λ -S component contains multiple doublets including $1^2\Sigma^-$. According to the electric dipole transition selection rule, the doublet-doublet transition is allowed, while the quadruplet-doublet transition is forbidden, so the TDM near the equilibrium internuclear distance from the 1/2(II) state to the ground state is much smaller than that from the other 1/2(III) and 3/2(III) states to the ground state.

The Franck-Condon factors for the transitions $2^2\Sigma^+$-$X^2\Pi$, $3^2\Sigma^+$-$X^2\Pi$, $1^2\Delta$-$X^2\Pi$, $2^2\Delta$-$X^2\Pi$ and 1/2(II)-$X^2\Pi_{3/2}$, 1/2(III)-$X^2\Pi_{3/2}$, 3/2(III)-$X^2\Pi_{3/2}$ were obtained by using the LEVEL program[26], and the related calculation results are listed in Table 5. The Franck-Condon factors of the above two transition states increase with the increase of the vibrational levels $v'$ and $v''$.

The lifetime of spontaneous emission is calculated by

$$\tau_{v'} = (A_{v'})^{-1} = \frac{3h}{64\pi^4 |a_0 \cdot e \cdot TDM|^2 \sum_{v''} q_{v',v''} (\Delta E_{v',v''})^3}$$
$$= \frac{4.936\times 10^5}{|TDM|^2 \sum_{v''} q_{v',v''} (\Delta E_{v',v''})^3} \quad (1)$$

Among them, $a_0$ represents the Bohr radius, $h$ is Planck's constant, $q_{v'v''}$ is the Franck-Condon factor between the $v'$ and $v''$ vibrational states, TDM is the transition dipole moment in a.u., and $\Delta E_{v'v''}$ (in $cm^{-1}$) is the energy separation of the vibrational level $v'$ from $v''$. The radiative lifetime $\tau_{v'}$ (in s) of the vibrational level with the lower bound state energy can be obtained by applying (1). The radiative lifetimes of the $2^2\Sigma^+$-$X^2\Pi$, $3^2\Sigma^+$-$X^2\Pi$, $1^2\Delta$-$X^2\Pi$, $2^2\Delta$-$X^2\Pi$ and 1/2(II)-$X^2\Pi_{3/2}$, 1/2(III)-$X^2\Pi_{3/2}$, 3/2(III)-$X^2\Pi_{3/2}$ transition states were listed in Table 6. The radiative lifetimes of $1^2\Delta$-$X^2\Pi$ and 1/2(II)-$X^2\Pi_{3/2}$ transitions decrease significantly with increasing vibrational quantum number. Because the TDM of 1/2(II)-$X^2\Pi_{3/2}$ near the equilibrium internuclear distance is much smaller than that of 1/2(III)-$X^2\Pi_{3/2}$, 3/2(III)-$X^2\Pi_{3/2}$ transition; This makes the vibrational level radiative lifetime of the 1/2(II) state significantly longer than that of the 1/2(III) and 3/2(III) states.

## 4. Conclusion

The electronic structures of 21 Λ-S states related to the two lowest dissociation limits of $ICl^+$ molecular ion are calculated by using the MRCI+Q method. The SOC effect and the core-valence electron correlation effect are considered in the calculation. The PECs of 21 Λ-S states and 42 Ω states were determined. Based on the calculated PECs, the spectroscopic constants of the bound state are calculated, which are in good agreement with the experimental results. Near the avoided crossing point, the abrupt changes of the DMs of $2^2\Sigma^+/3^2\Sigma^+$ and $2^2\Pi/3^2\Pi$ states are caused by the changes of the configuration components of the corresponding states. The complex interactions between $2^2\Pi$, $3^2\Pi$, $1^2\Delta$, $2^2\Delta$ and nearby states are elucidated with the help of theoretically calculated SOC matrix elements. The $v'$=0 vibrational level of the $2^2\Pi$ state predissociate through the $1^2\Delta$ state, the $v'$=0 vibrational level of the $1^2\Delta$ state predissociate through the $1^4\Delta$ state, and the $v'$=0 vibrational levels of the $2^2\Delta$ state and the $3^2\Pi$ state predissociate through the $2^4\Pi$ and $1^4\Pi$ states, respectively. In addition, the TDMs, Franck-Condon Factors, and radiative lifetimes for the $2^2\Sigma^+$-$X^2\Pi$, $3^2\Sigma^+$-$X^2\Pi$, $1^2\Delta$-$X^2\Pi$, $2^2\Delta$-$X^2\Pi$ and 1/2(II)-$X^2\Pi_{3/2}$, 1/2(III)-$X^2\Pi_{3/2}$, 3/2(III)-$X^2\Pi_{3/2}$ transitions were calculated.

**Data Availability Statement**

The data sets supporting this research are available in the supplementary materials (https://wulixb.iphy.ac.cn/cn/article/doi/10.7498/aps.74.20250510) and the Science Data Bankhttps://doi.org/10.57760/sciencedb.j00213.00140.

**Table 1.** Electronic configuration and vertical excitation energies of $ICl^+$ molecular ion at $R_e$.

| Λ-S States | Main configuration at respective $R_e$ (%) | $T$/cm$^{-1}$ |
|---|---|---|
| $X^2\Pi$ | $7\sigma^2 8\sigma^2 9\sigma^2 10\sigma^2 11\sigma^0 3\pi^4 4\pi^4 5\pi^3$(91) | 0 |
| $1^4\Sigma^-$ | $7\sigma^2 8\sigma^2 9\sigma^2 10\sigma^2 11\sigma^1 3\pi^4 4\pi^4 5\pi^2$(77) | 18885.3 |
| | $7\sigma^2 8\sigma^2 9\sigma^2 10\sigma^2 11\sigma^1 3\pi^4 4\pi^3 5\pi^3$(20) | |
| $2^2\Pi$ | $7\sigma^2 8\sigma^2 9\sigma^2 10\sigma^2 11\sigma^0 3\pi^4 4\pi^3 5\pi^4$(87) | 23660.8 |
| | $7\sigma^2 8\sigma^2 9\sigma^2 10\sigma^1 11\sigma^1 3\pi^4 4\pi^4 5\pi^3$(6) | |
| $1^2\Delta$ | $7\sigma^2 8\sigma^2 9\sigma^2 10\sigma^2 11\sigma^1 3\pi^4 4\pi^4 5\pi^2$(67) | 25549.0 |
| | $7\sigma^2 8\sigma^2 9\sigma^2 10\sigma^2 11\sigma^1 3\pi^4 4\pi^3 5\pi^3$(29) | |
| $1^2\Sigma^+$ | $7\sigma^2 8\sigma^2 9\sigma^2 10\sigma^2 11\sigma^1 3\pi^4 4\pi^4 5\pi^2$(59) | 28700.8 |
| | $7\sigma^2 8\sigma^2 9\sigma^2 10\sigma^2 11\sigma^1 3\pi^4 4\pi^3 5\pi^3$(34) | |
| $2^2\Sigma^+$ | $7\sigma^2 8\sigma^2 9\sigma^2 10\sigma^1 11\sigma^0 3\pi^4 4\pi^4 5\pi^4$(94) | 33963.0 |
| $1^4\Delta$ | $7\sigma^2 8\sigma^2 9\sigma^2 10\sigma^2 11\sigma^1 3\pi^4 4\pi^3 5\pi^3$(99) | 35589.7 |
| $2^2\Delta$ | $7\sigma^2 8\sigma^2 9\sigma^2 10\sigma^2 11\sigma^1 3\pi^4 4\pi^3 5\pi^3$(96) | 43617.4 |
| $3^2\Sigma^+$ | $7\sigma^2 8\sigma^2 9\sigma^2 10\sigma^2 11\sigma^1 3\pi^4 4\pi^3 5\pi^3$(95) | 44332.0 |
| $3^2\Sigma^-$ | $7\sigma^2 8\sigma^2 9\sigma^2 10\sigma^2 11\sigma^1 3\pi^4 4\pi^3 5\pi^3$(67) | 52169.3 |
| | $7\sigma^2 8\sigma^2 9\sigma^2 10\sigma^1 11\sigma^2 3\pi^4 4\pi^4 5\pi^2$(22) | |
| | $7\sigma^2 8\sigma^2 9\sigma^2 10\sigma^2 11\sigma^1 3\pi^4 4\pi^4 5\pi^2$(9) | |
| $3^2\Pi$ | $7\sigma^2 8\sigma^2 9\sigma^2 10\sigma^2 11\sigma^2 3\pi^4 4\pi^4 5\pi^1$(48) | 53534.0 |
| | $7\sigma^2 8\sigma^2 9\sigma^2 10\sigma^2 11\sigma^2 3\pi^4 4\pi^3 5\pi^2$(48) | |

**Table 2.** Spectroscopic constants of the Λ-S states of $ICl^+$ molecular ion.

| Λ-S States | | $T_e$/cm$^{-1}$ | $D_0$/eV | $B_e$/cm$^{-1}$ | $\omega_e$/cm$^{-1}$ | $R_e$/Å |
|---|---|---|---|---|---|---|
| $X^2\Pi$ | This Work. | 0 | 2.50 | 0.1221 | 432 | 2.24 |
| | Expt.[a)] | 0 | 2.52 | | | |
| $2^2\Pi$ | This Work. | 18218 | 0.27 | 0.0915 | 259 | 2.59 |
| | Expt.[b)] | 19551 | | | | |
| | Calc.[c)] | 14352 | | | 207 | 2.78 |
| $1^2\Delta$ | This Work. | 17516 | 0.35 | 0.0737 | 109 | 2.89 |
| $1^4\Sigma^-$ | This Work. | 13617 | 0.83 | 0.0889 | 218 | 2.63 |
| $1^4\Delta$ | This Work. | 17183 | 0.39 | 0.0670 | 138 | 3.03 |
| $1^2\Sigma^+$ | This Work. | 18015 | 0.28 | 0.0654 | 153 | 3.05 |
| $3^2\Pi$ | This Work. | 22122 | 0.80 | 0.0628 | 526 | 3.14 |
| | Expt.[a)] | 22420 | | | | |
| $2^2\Delta$ | This Work. | 24768 | 0.52 | 0.0696 | 171 | 2.97 |
| $2^2\Sigma^+$ | This Work. | 25192 | 0.47 | 0.0687 | 164 | 2.99 |
| $3^2\Sigma^-$ | This Work. | 25241 | 0.46 | 0.0630 | 181 | 3.13 |

Ref[a][33], Ref[b][12], Ref[c][16].

.

**Table 3.** Dissociation relationships of Ω states of $ICl^+$ molecular ion.

| Atomic state ($I^+$+Cl) | Ω states | Energy($cm^{-1}$) This work | Energy($cm^{-1}$) Expt.[a] |
|---|---|---|---|
| $I^+(^3Pg_2)+Cl(^2Pu_{3/2})$ | 7/2, 5/2(2), 3/2(3), 1/2(4) | 0 | 0 |
| $I^+(^3Pg_2)+Cl(^2Pu_{1/2})$ | 5/2, 3/2(2), 1/2(2) | 821 | 882 |
| $I^+(^3Pg_0)+Cl(^2Pu_{3/2})$ | 3/2, 1/2 | 6433 | 6448 |
| $I^+(^3Pg_1)+Cl(^2Pu_{3/2})$ | 5/2, 3/2(2), 1/2(3) | 6901 | 7087 |
| $I^+(^3Pg_0)+Cl(^2Pu_{1/2})$ | 1/2 | 7254 | 7330 |
| $I^+(^3Pg_1)+Cl(^2Pu_{1/2})$ | 3/2, 1/2(2) | 7722 | 7969 |
| $I^+(^1Dg_2)+Cl(^2Pu_{3/2})$ | 7/2, 5/2(2), 3/2(3), 1/2(4) | 13950 | 13727 |
| $I^+(^1Dg_2)+Cl(^2Pu_{1/2})$ | 5/2, 3/2(2), 1/2(2) | 14771 | 14610 |

Ref[a][34].

**Table 4.** Spectroscopic constants of Ω states of $ICl^+$ molecular ion.

| Ω States | | $T_e$/cm$^{-1}$ | $D_0$/eV | $B_e$/cm$^{-1}$ | $\omega_e$/cm$^{-1}$ | $R_e$/Å | main Λ-S composition at $R_e$(%) |
|---|---|---|---|---|---|---|---|
| $X^2\Pi_{3/2}$ | This work | 0 | 2.31 | 0.1216 | 419 | 2.250 | $X^2\Pi$ (98.1) |
| | EXP.[a)] | | | | 390 | | |
| | EXP.[b)] | | | | 429 | 2.240±0.01 | |
| | Cal.[c)] | | | | 311 | 2.470 | |
| $X^2\Pi_{1/2}$ | This work | 4501 | 1.75 | 0.1217 | 426 | 2.248 | $X^2\Pi$ (97.7) |
| | EXP.[d)] | 4680 | | | | | |
| | EXP.[b)] | 4670±16 | | | 437 | 2.224±0.001 | |
| | Cal.[c)] | 5424 | | | 314 | 2.460 | |
| 1/2(II) | This work | 14808 | 0.50 | 0.0864 | 244 | 2.666 | $1^4\Sigma^-$(79.7)<br>$1^2\Sigma^+$ (12.5) |
| 1/2(III) | This work | 17191 | 0.21 | 0.0656 | 78 | 3.063 | $1^2\Sigma^-$ (38.3)<br>$1^4\Sigma^+$ (25.7)<br>$3^2\Pi$ (10.8)<br>$1^2\Pi$ (5.8) |
| 3/2(II) | This work | 15202 | 0.44 | 0.0831 | 156 | 1.518 | $1^4\Sigma^-$(80.7)<br>$1^4\Sigma^+$ (6.7) |
| 3/2(III) | This work | 16795 | 0.21 | 0.0649 | 107 | 3.003 | $1^4\Delta$ (58.3)<br>$1^2\Delta$ (34.8) |

Ref[a][6], Ref[b][12], Ref[c][16], Ref[d][33].

**Table 5.** Frank-Condon factors for the transition of $ICl^+$ molecular ion.

| | | $v''$=0 | $v''$=1 | $v''$=2 | $v''$=3 | $v''$=4 |
|---|---|---|---|---|---|---|
| | | | $1^2\Delta$-$X^2\Pi$ | | | |
| $v'$=0 | ThisWork | $1.888\times10^{-17}$ | $8.612\times10^{-16}$ | $1.956\times10^{-14}$ | $2.919\times10^{-13}$ | $3.193\times10^{-12}$ |
| $v'$=1 | ThisWork | $3.271\times10^{-16}$ | $1.436\times10^{-14}$ | $3.138\times10^{-13}$ | $4.501\times10^{-12}$ | $4.724\times10^{-11}$ |
| $v'$=2 | ThisWork | $2.958\times10^{-15}$ | $1.252\times10^{-13}$ | $2.635\times10^{-12}$ | $3.635\times10^{-11}$ | $3.662\times10^{-10}$ |
| | | | $2^2\Delta$-$X^2\Pi$ | | | |
| $v'$=0 | ThisWork | $4.361\times10^{-27}$ | $5.255\times10^{-25}$ | $2.996\times10^{-23}$ | $1.085\times10^{-21}$ | $2.796\times10^{-20}$ |
| $v'$=1 | ThisWork | $1.373\times10^{-25}$ | $1.547\times10^{-23}$ | $8.325\times10^{-22}$ | $2.864\times10^{-20}$ | $7.046\times10^{-19}$ |
| $v'$=2 | ThisWork | $2.081\times10^{-24}$ | $2.230\times10^{-22}$ | $1.149\times10^{-20}$ | $3.797\times10^{-19}$ | $8.989\times10^{-18}$ |
| | | | $2^2\Sigma^+$-$X^2\Pi$ | | | |
| $v'$=0 | ThisWork | $5.553\times10^{-27}$ | $5.602\times10^{-25}$ | $2.715\times10^{-23}$ | $8.439\times10^{-22}$ | $1.889\times10^{-20}$ |
| $v'$=1 | ThisWork | $2.304\times10^{-25}$ | $2.065\times10^{-23}$ | $9.017\times10^{-22}$ | $2.559\times10^{-20}$ | $5.290\times10^{-19}$ |
| | | | $3^2\Sigma^+$-$X^2\Pi$ | | | |
| $v'$=0 | ThisWork | $4.161\times10^{-18}$ | $2.943\times10^{-16}$ | $1.010\times10^{-14}$ | $2.216\times10^{-13}$ | $3.456\times10^{-12}$ |
| $v'$=1 | ThisWork | $4.071\times10^{-17}$ | $2.814\times10^{-15}$ | $9.431\times10^{-14}$ | $2.019\times10^{-12}$ | $3.067\times10^{-11}$ |
| $v'$=2 | ThisWork | $2.033\times10^{-16}$ | $1.374\times10^{-14}$ | $4.503\times10^{-13}$ | $9.412\times10^{-12}$ | $1.394\times10^{-10}$ |
| | | | 1/2(II)-$X^2\Pi_{3/2}$ | | | |
| $v'$=0 | ThisWork | $1.226\times10^{-9}$ | $3.102\times10^{-8}$ | $3.870\times10^{-7}$ | $3.155\times10^{-6}$ | $1.872\times10^{-5}$ |
| $v'$=1 | ThisWork | $1.923\times10^{-8}$ | $4.431\times10^{-7}$ | $4.996\times10^{-6}$ | $3.649\times10^{-5}$ | $1.921\times10^{-4}$ |
| $v'$=2 | ThisWork | $1.679\times10^{-7}$ | $3.501\times10^{-6}$ | $3.541\times10^{-5}$ | $2.296\times10^{-4}$ | $1.060\times10^{-3}$ |
| | | | 1/2(III)-$X^2\Pi_{3/2}$ | | | |
| $v'$=0 | ThisWork | $1.314\times10^{-22}$ | $9.703\times10^{-21}$ | $3.517\times10^{-19}$ | $8.368\times10^{-18}$ | $1.457\times10^{-16}$ |
| $v'$=1 | ThisWork | $5.532\times10^{-21}$ | $3.879\times10^{-19}$ | $1.346\times10^{-17}$ | $3.083\times10^{-16}$ | $5.184\times10^{-15}$ |
| $v'$=2 | ThisWork | $2.944\times10^{-20}$ | $1.987\times10^{-18}$ | $6.647\times10^{-17}$ | $1.468\times10^{-15}$ | $2.379\times10^{-14}$ |
| | | | 3/2(III)-$X^2\Pi_{3/2}$ | | | |
| $v'$=0 | ThisWork | $1.106\times10^{-22}$ | $5.929\times10^{-21}$ | $1.542\times10^{-19}$ | $2.607\times10^{-18}$ | $3.216\times10^{-17}$ |
| $v'$=1 | ThisWork | $3.736\times10^{-21}$ | $1.836\times10^{-19}$ | $4.435\times10^{-18}$ | $7.052\times10^{-17}$ | $8.256\times10^{-16}$ |

**Table 6.** Radiative lifetimes of $ICl^+$ molecular ion.

| Transition | | Radiative lifetimes (s) | | |
|---|---|---|---|---|
| | | $v'$=0 | $v'$=1 | $v'$=2 |
| $1^2\Delta$-$X^2\Pi$ | ThisWork | $5.69\times10^{-3}$ | $4.92\times10^{-3}$ | $4.40\times10^{-3}$ |
| $2^2\Delta$-$X^2\Pi$ | ThisWork | $5.83\times10^{-5}$ | $6.18\times10^{-5}$ | $6.99\times10^{-5}$ |
| $2^2\Sigma^+$-$X^2\Pi$ | ThisWork | $2.91\times10^{-5}$ | $3.28\times10^{-5}$ | |
| $3^2\Sigma^+$-$X^2\Pi$ | ThisWork | $1.58\times10^{-5}$ | $1.66\times10^{-5}$ | $1.83\times10^{-5}$ |
| 1/2(II)-$X^2\Pi_{3/2}$ | ThisWork | $3.73\times10^{-2}$ | $3.42\times10^{-2}$ | $3.09\times10^{-2}$ |
| 1/2(III)-$X^2\Pi_{3/2}$ | ThisWork | $1.38\times10^{-2}$ | $5.43\times10^{-3}$ | $7.61\times10^{-3}$ |
| 3/2(III)-$X^2\Pi_{3/2}$ | ThisWork | $3.28\times10^{-2}$ | $2.06\times10^{-2}$ | |

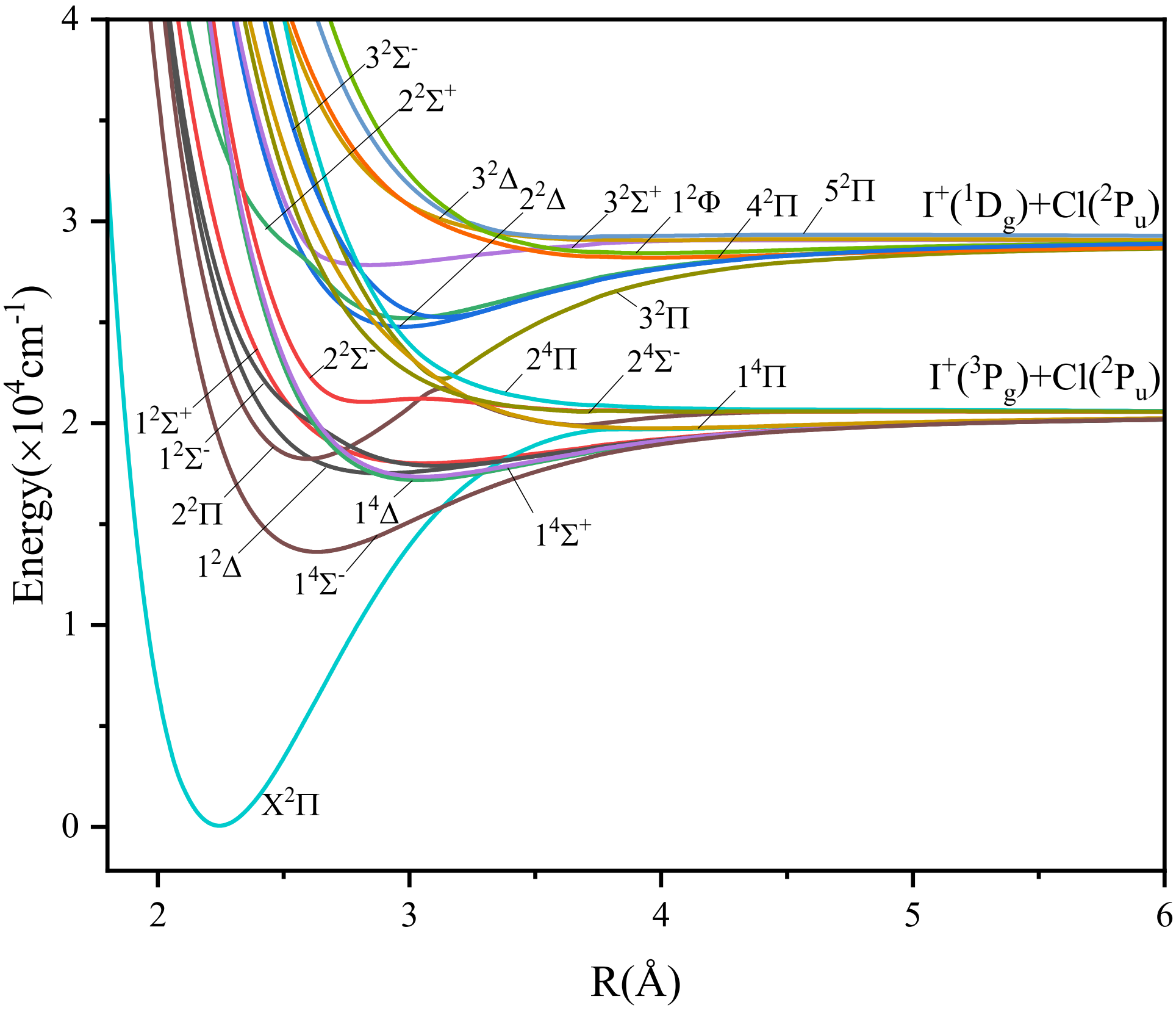

**Figure 1.** Potential energy curves of the Λ-S states of $ICl^+$ molecular ion.

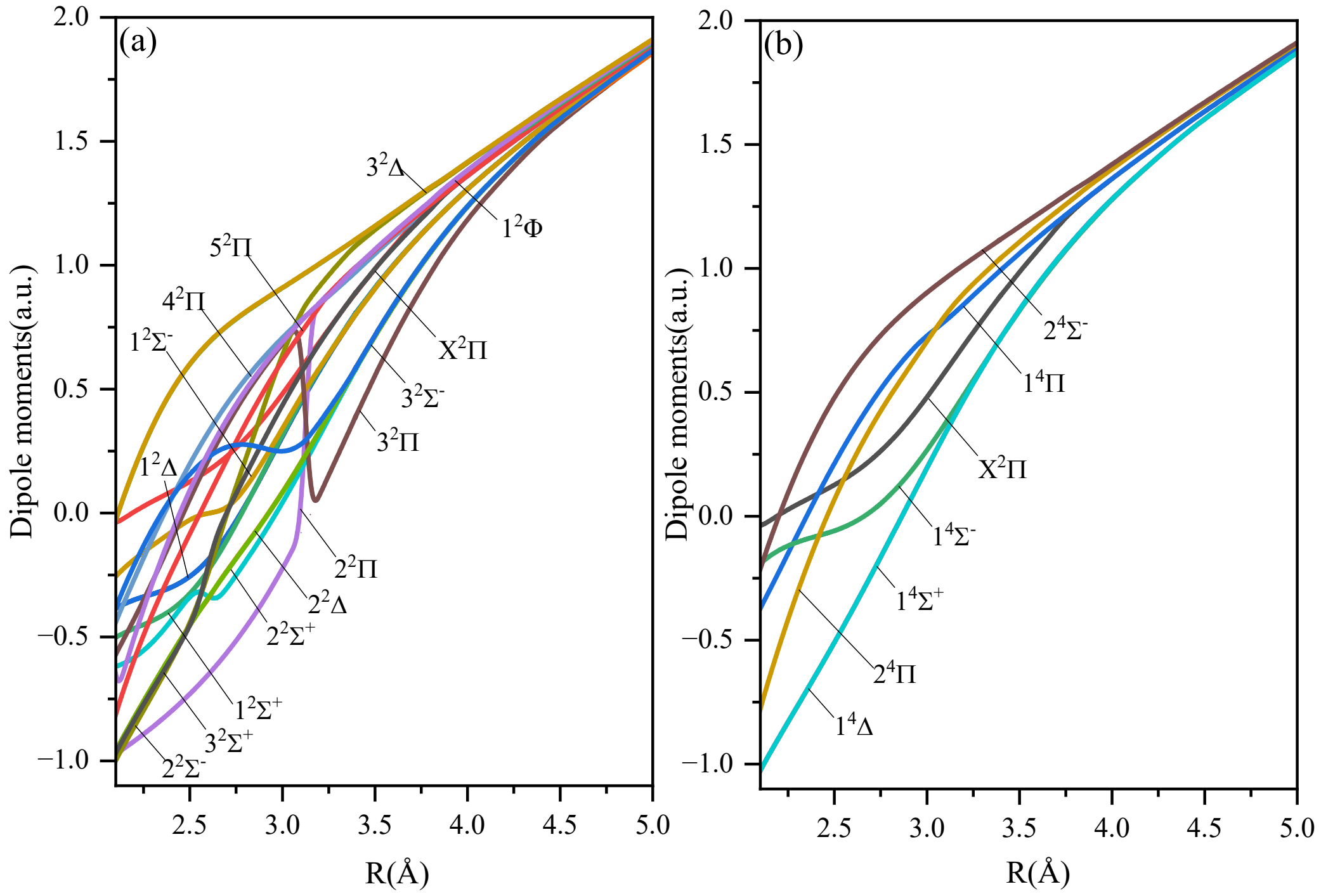

**Figure 2.** Dipole moments curves of the Λ-S states of $ICl^+$ molecular ion: (a) dipole moment of the doublet state; (b) dipole moment of the quartet state.

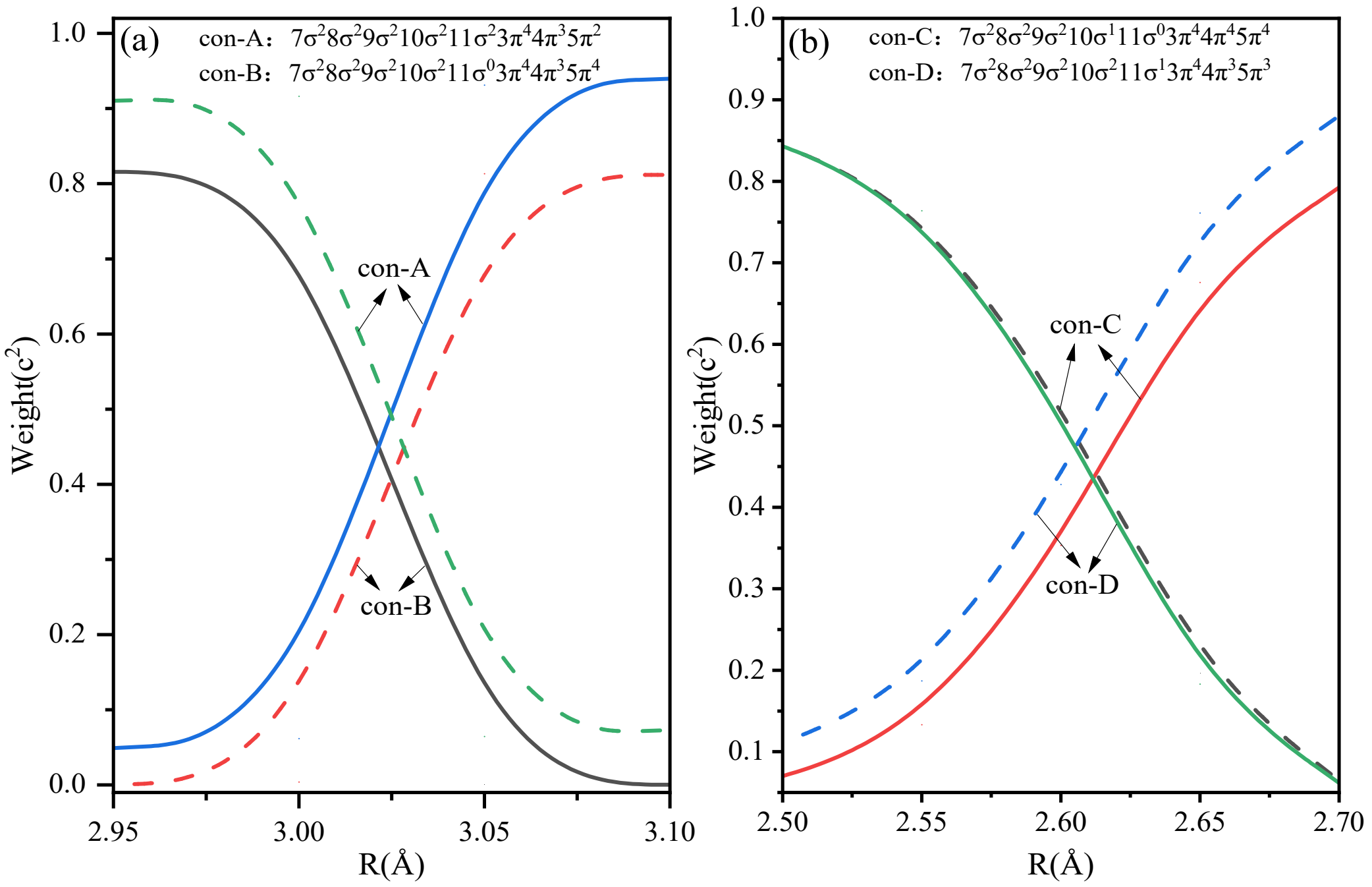

**Figure 3.** The *R*-dependent weights ($c^2$) of the electronic configurations of $2^2\Pi/3^2\Pi$ (a) and $2^2\Sigma^+/3^2\Sigma^+$ (b) states.

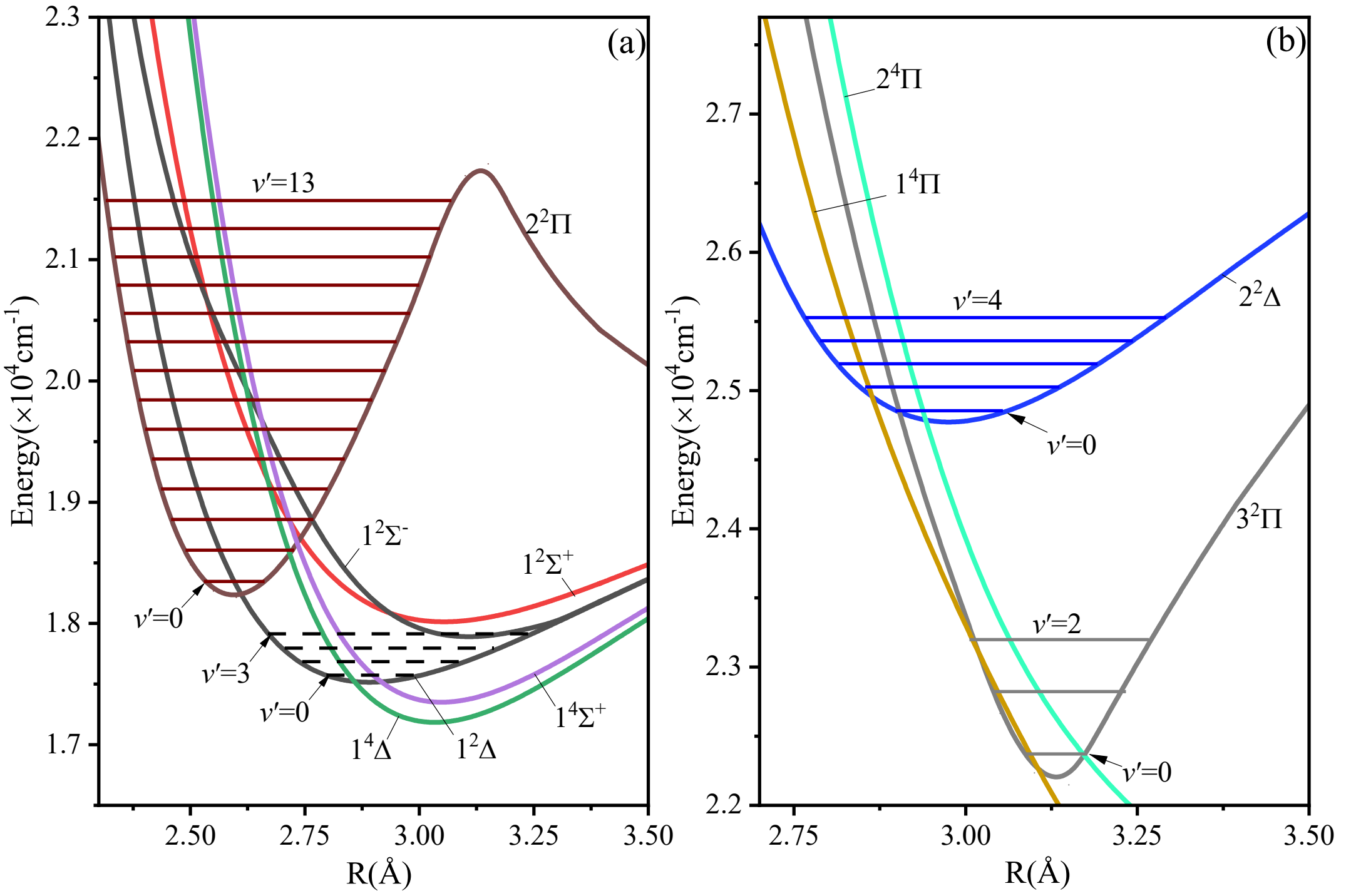

**Figure 4.** An amplified view of crossing region with the corresponding vibrational levels.

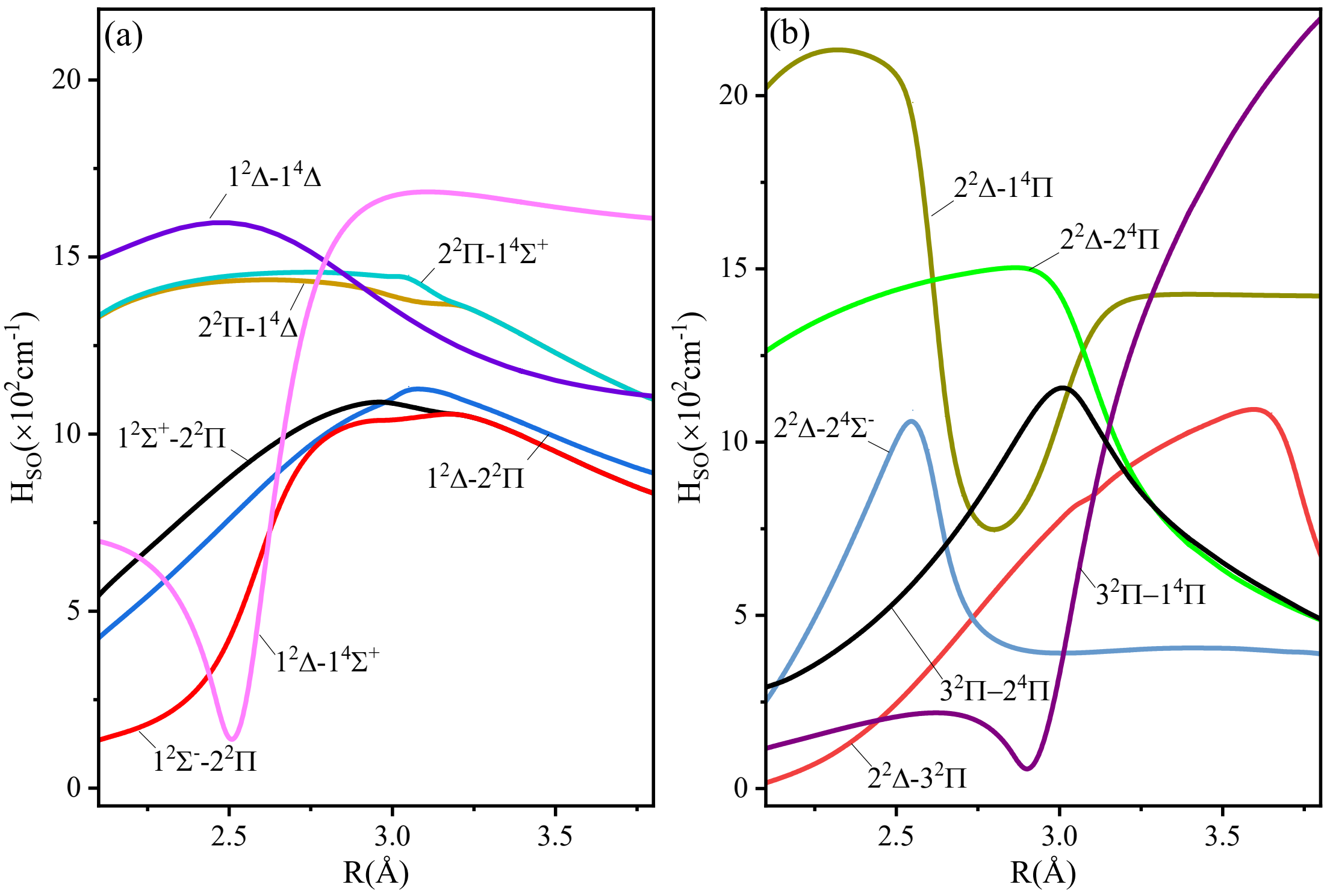

**Figure 5.** Variation curves of spin-orbit coupling matrix elements of $2^2\Pi$, $1^2\Delta$ states (a) and $2^2\Delta$, $3^2\Pi$ states (b) with internuclear distance.

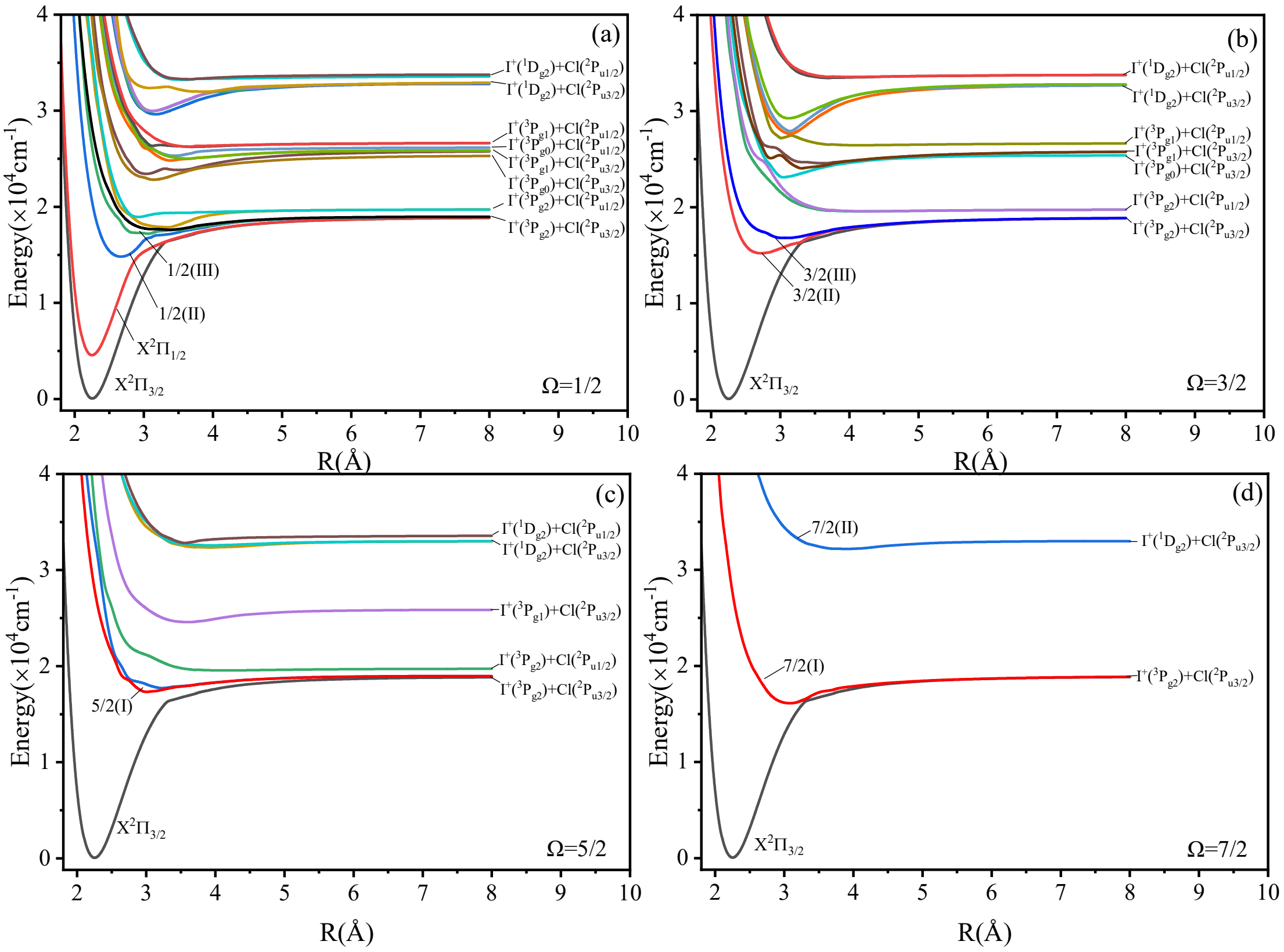

**Figure 6.** Potential energy curves of the Ω states of $ICl^+$ molecular ion.

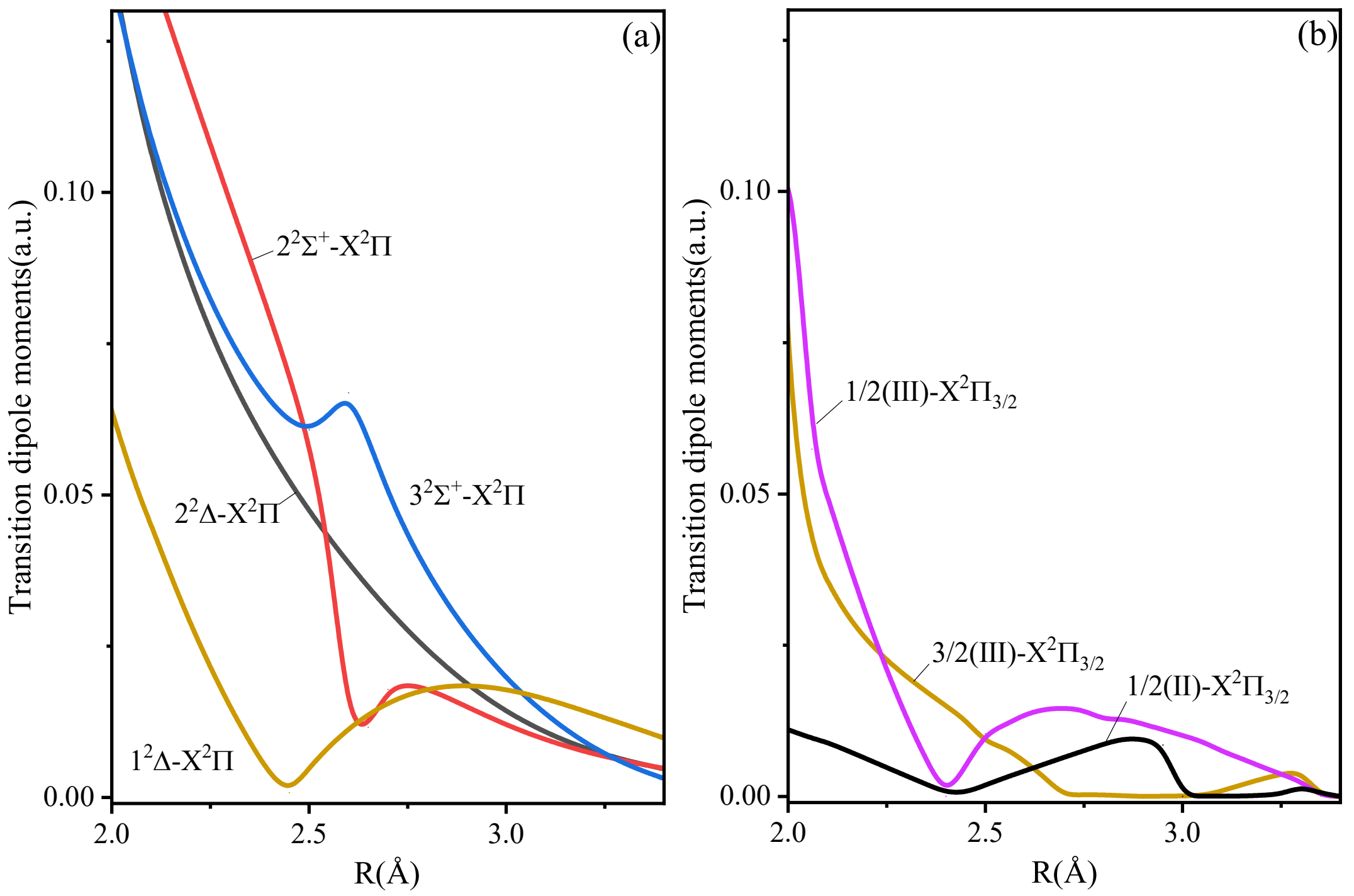

**Figure 7.** Transition dipole moments curves of $ICl^+$ molecular ion.